# Excess Volume at Grain Boundaries in *hcp* Metals

J. L. Cao and W. T. Geng[*]

*School of Materials Science & Engineering, University of Science and Technology Beijing, Beijing 100083, China*

**Abstract**

The excess volume associated with grain boundaries represents a key structural parameter for the characterization of grain boundaries. It plays a critical role in segregation of impurity and alloy elements to grain boundaries, and influences significantly the mechanical and functional properties of materials. We have carried out first-principles density functional calculations on the atomic structure of the $(10\bar{1}2)$ coherent twin boundary in hexagonal close packed (*hcp*) Ti and Zr and the basal-prismatic boundary in Zn, Cd, and Zr. We find the calculated excess volume has a vanishing magnitude at the $(10\bar{1}2)$ coherent twin boundary in Ti and Zr; whereas it is remarkable at the basal-prismatic boundary in Zn, Cd, and Zr.

---

[*]E-mail: geng@ustb.edu.cn

The excess volume of a grain boundary (GB) represents a key structural parameter for the characterization of GBs. Over half a century ago, Seeger and Schotty, (1) postulated that the excess energy of high-angle GBs in silver and copper was linearly interrelated with the excess volume. With more sophisticated electronic structure calculations, Ferrante and Smith (2) demonstrated that the normalized expansion parameter was virtually the key parameter in the energies of bimetallic interfaces, and the energy as a function of excess volume was insensitive to the detailed atomic structure. Even earlier, Read and Shockley (3) developed a dislocation model to describe the GBs by the arrangement of edge dislocations. Wolf formulated the relation of GB energy and excess volume (4) as a function of the mis-orientation angle between the adjoining grains and the Burgers vector. In all these models, it was assumed *a priori* that GBs possessed positive excess volume to accommodate the structure mismatch with lower coordination and extended average bond length. Generally, the GBs with the smallest volume expansion were the most stable ones and the most prone to present in real materials.

Over the years, through intensive researches scientists have made great progress in measuring with high precision the excess volume of GBs. Rotating-sphere-on-a-plane measurements suggested the GBs with the minimum energy bear smaller excess volume than higher-energy GBs. (5) High-resolution transmission electron microscopy technique can provide direct evidence for the existence of volume expansion at GBs. (6, 7) Other approaches to determine the excess volume of GBs include a thermodynamic method employing the pressure dependence of GB energy in Al tri-crystals, (8) density measurement (9) and grain growth kinetics modeling (10) for nano-crystalline materials.

In a recent DFT computational work, Kumar *et al.* investigated the energy and atomic solubility of twining-associated boundaries in hexagonal metals Be, Mg, Ti, Zr, Zn, and Cd (11). It was reported in this work that the excess volume of the $(10\bar{1}2)$ coherent twin boundary (CTB) in Ti and Zr is negative, $e_{GB}$=-0.03 Å and $e_{GB}$=-0.04 Å, respectively. Moreover, the excess volume of the coherent basal-prismatic boundary (CBP) in Zn ($e_{GB}$=-0.08 Å) and Zr ($e_{GB}$=-0.04 Å) is also noticeably negative. Although these numerical results were not the key points the authors would emphasize, we feel they deserve a more rigorous scrutiny for this computational result is no doubt very much unexpected. Thus, we are strongly motivated to revisit the GBs with negative excess volume.

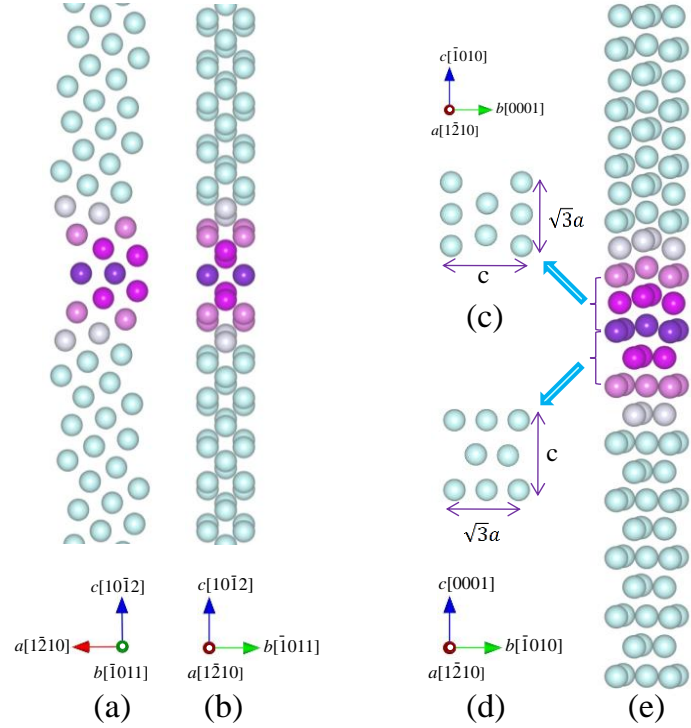

**Fig. 1. Side views of the computation cells used to model the $(10\bar{1}2)$ coherent twin boundary(a, b) and coherent basal-prismatic boundary (c-e) in *hcp* metals. Crystal orientations and unit cells of the basal-crystal and prismatic-crystal are sketched in (c) and (d).**

The first-principles DFT calculations were performed using the Vienna *Ab initio* Simulation Package (VASP) (12). The interaction between values electron and nuclei was obtained using projector augmented wave (PAW) method (13). The exchange and correlation terms in DFT method were treated with generalized gradient approximation (GGA) in the Perdew-Burke-Ernzerhof (PBE) form (14). A conjugate gradient algorithm was used to relax the atomic positions to a local minimum in the total energy landscape. The cut-off energy for the plan wave basic set was set to 480 eV. For each system, geometry optimization would continue until the total energy of this system was converged to less than $10^{-4}$ eV. An 8×8×2 *k*-mesh in Monkhorst-Pack Scheme replaces the integration over Brillouin zone in the following simulation.

The free surface (FS) and GB are simulated by slabs. As sketched in Fig. 1(a) and 1(b), the $(10\bar{1}2)$ coherent twin boundary in an *hcp* metal, a symmetric tilt GB, can be created by stacking the mirror image of a FS slab $[10\bar{1}2]$ crystal orientation. The *a* axis of this GB supercell is along the zonal direction $[1\bar{2}10]$, and the *b* axis is along the $[\bar{1}011]$ direction. The structure within the lateral plane was kept unchanged to maintain in-plane symmetry and the equilibrium atomic positions in the vertical direction of the FS and GB were fully relaxed without additional stresses. The FS system was created to serve as reference state in the following calculations. To obtain a reliable energy difference, the FS and GB systems must be treated on an equal footing. Therefore, the same set of numerical parameters in DFT calculations for both GB and FS were used. An 8×8×2 *k*-mesh in Monkhorst-Pack Scheme replaces the integration over Brillouin zone in the following simulation. As shown in Figs. 1(c) and 1(d), we give crystal orientations and unit cells of the basal-crystal and prismatic-crystal, and Fig. 1(e) is the atomic structure of a relaxed coherent basal-prismatic boundary (CTB). Different from Ref. 11, where the lattice constant of the lateral plane is the average of that of the two surfaces, i.e., $y = (\sqrt{3}a + c)/2$, we perform series of calculations by varying the lattice constant to find the optimized value.

Prior to carrying out the computation, we had already realized that the slabs Kumar *et al.* adopted to simulate the GBs might not be thick enough. Each of the two grains in CTB and CBP slabs contains only twenty atomic layers. In our slab model, we separate the neighboring slabs in $[10\bar{1}2]$ direction by a vacuum region of at least 10 Å to minimize the interaction between slabs. The excess volumes (per unit area) of CTBs are described as:

$$e_{GB} = (V_{GB} - V_{FS})/A = d_{GB} - d_{FS} \qquad (1),$$

where $V_{GB}$ and $V_{FS}$ are the volume of periodic atomic units of GB and FS, respectively, $A$ is the area of grain boundary plane, $d_{FS}$, $d_{GB}$ are the inter-layer distance between the top and the bottom layer in the unit cell at FS and GB. For CTB of Ti, we find that with 21-layer grains, $e_{GB}$=-0.058Å; with 29-layer grains, $e_{GB}$=+0.001Å; with 37-layer grains, $e_{GB}$=+0.009Å; and with 45-layer grains, $e_{GB}$=-0.004Å. And for Zr, $e_{GB}$=-0.075Å, -0.070Å, -0.020Å, and -0.004Å, respectively. Thus, our calculations demonstrate unambiguously that if the slabs are thick-enough to yield a bulk-like environment in the center of the grain, or equally put, to minimize the GB-GB interactions, there would appear no negative excess volume for the CTB of Ti and Zr, the results are listed in Table I.

**Table I Calculated excess volumes (Å) at coherent twin boundary (CTB) in Ti, Zr and coherent basal-prismatic boundary (CBP) in Zn, Zr and Cd.**

|  | CTB | | CBP | | |
| --- | --- | --- | --- | --- | --- |
|  | Ti | Zr | Zn | Zr | Cd |
| $e_{GB}$ (Å) | 0.00 | 0.00 | +0.20 | +0.38 | +0.10 |

As for the CBP boundaries of the *hcp* metals, we find a 24-layer slab is appropriate to model the GB, and a 20-layer slab, as adopted in the original work, is also roughly thick enough. However, we noticed that the treatment of lattice mismatch between basal and prismatic surfaces across the CBP was not without question. To make a CBP interface, it was assumed in Ref. 11 that the final lattice constant after adjustment will be the average of those of two surfaces, i.e., $y = (\sqrt{3}a + c)/2$. This assumption, unfortunately, is not well founded. The reason is that the basal plane is a densely-packed atomic plane, whereas the prismatic plane is rather loosely-packed. As a consequence, their compressibility must be very different. We would expect the relaxed CBP interface system has two-dimensional lattice constants closer to those of the basal plane (Zn and Zr) or prismatic plane (Zr). The compression experienced by the prismatic or basal surface slab will lead an enlarged thickness in the normal direction. The excess volumes (per unit area) of CBPs are described as:

$$e_{GB} = (V_{GB}' - V_{FS}')/A \qquad (2),$$

where $V_{GB}'$ and $V_{FS}'$ are the volume of the central zone (about half of the total units to removing the effect of free surface) in periodic atomic units of GB and FS, respectively, $A$ is the area of grain boundary plane.

Our DFT calculations confirms this rationale and show that for Zn, the optimized $y$ is 1.74$a$, only slightly larger than the value for the basal plane, $y$=1.732$a$, significantly smaller than that for the prismatic plane, $y$=$c$=1.82$a$. More importantly, the excess volume is positive, $e_{GB}$=+0.10Å. For CBP interface in Zr and Cd, we find $e_{GB}$=+0.20Å, $e_{GB}$=+0.38Å, as shown in Table I.

In conclusion, through careful examination of the thickness of the slabs used to model the twin boundary and the lateral lattice constants of the basal-prismatic boundary, our rigorous calculations demonstrate that the excess volume has a vanishing magnitude at the ($10\bar{1}2$) coherent twin boundary in Ti and Zr. By comparison, the excess volume is remarkably positive at the basal-prismatic boundary in Zn, Cd, and Zr.